\documentclass[aps,prc,twocolumn,floatfix,superscriptaddress,amsmath,amssymb,nofootinbib]{revtex4-2}

\usepackage[utf8]{inputenc} 
\usepackage[T1]{fontenc}
\usepackage{dcolumn}
\usepackage{braket}
\usepackage{float}
\usepackage{graphicx}
\usepackage[colorlinks=true,linkcolor=blue,citecolor=blue,urlcolor=blue]{hyperref}
\usepackage{orcidlink}
\usepackage{textcase}

\graphicspath{{.}{./figures/}}

\begin{document}

\title{Absence of a shell closure in $^{140}$Sn}

\author{Francesca~Bonaiti\orcidlink{0000-0002-3926-1609}}
\email{bonaiti@frib.msu.edu}
\affiliation{Facility for Rare Isotope Beams, Michigan State University, East Lansing, Michigan 48824, USA}
\affiliation{Physics Division, Oak Ridge National Laboratory, Oak Ridge, Tennessee 37831, USA}

\author{Bingcheng~He\orcidlink{0000-0003-4483-1992}}
\email{bhe9@utk.edu}
\affiliation{Department of Physics and Astronomy, University of Tennessee, Knoxville, Tennessee 37996, USA}

\author{Gaute~Hagen\orcidlink{0000-0001-6019-1687}}
\affiliation{Physics Division, Oak Ridge National Laboratory, Oak Ridge, Tennessee 37831, USA}
\affiliation{Department of Physics and Astronomy, University of Tennessee, Knoxville, Tennessee 37996, USA}

\author{Thomas~Papenbrock\orcidlink{0000-0001-8733-2849}}
\affiliation{Department of Physics and Astronomy, University of Tennessee, Knoxville, Tennessee 37996, USA}
\affiliation{Physics Division, Oak Ridge National Laboratory, Oak Ridge, Tennessee 37831, USA}

\begin{abstract}
There are conflicting theoretical results about the presence of a shell closure in the neutron-rich nucleus $^{140}$Sn. We address this controversy by performing \textit{ab initio} computations, using a nuclear interaction from chiral effective field theory that accurately reproduced and predicted low-lying states in doubly magic nuclei. We verify that this interaction accurately reproduces low-lying states in $^{133}$Sn. We assume that $^{140}$Sn exhibits a closed $7/2^-$ neutron subshell beyond $^{132}$Sn and compute its first excited $2^+$ state. The resulting energy is small and this contradicts the assumption.  
\end{abstract}
\maketitle

\section{Introduction}
\label{intro}
One of the most interesting questions in nuclear structure is understanding how shell closures evolve in exotic neutron-rich nuclei far from stability. While the canonical magic numbers govern stable nuclei, both the disappearance of established shell closures and the emergence of new ones have been documented in neutron-rich systems~\cite{otsuka2020,nowacki2021}. In medium-mass nuclei, this is shown for instance by the emergence of unconventional subshell closures in oxygen and calcium isotopes at $N = 14, 16$ \cite{thirolf2000,kanungo2009,otsuka2010} and $N = 32, 34$ \cite{steppenbeck2013,steppenbeck2013b}, respectively. Moving to heavier systems, the evolution of shell structure beyond the doubly-magic nucleus $^{132}$Sn has attracted considerable theoretical and experimental attention~\cite{bjornstad1980,jones2010,jin2011,allmond2014,naidja2017}. This region is of astrophysical relevance: the $A\approx130$ peak in the r-process abundance pattern is tied to nuclear structure properties near the $N = 82$ shell closure~\cite{mumpower2016}. 

Based on trends of $2^+$ excitation energies from an empirical shell-model interaction beyond $N = 82$, \textcite{sarkar2010} suggested that the $N = 90$ nucleus $^{140}$Sn exhibits a shell closure. This prediction has since then stimulated spectroscopic studies~\cite{simpson2014,lozeva2016,jungclaus2024,lozeva2024} and mean-field and shell-model calculations~\cite{covello2011,jin2011,anguiano2012,maheshwari2015,naidja2015,naidja2017} in neutron-rich nuclei beyond $^{132}$Sn. Shell-model calculations based on more realistic effective interactions predict little variation in the first $2^+$ excitation energy between $^{132}$Sn and $^{140}$Sn~\cite{covello2011,naidja2015}, in contrast to the pronounced rise expected at a shell closure. More recently, a measurement of the $3/2^-$ single-particle state in $^{135}$Sn found good agreement with a shell-model interaction disfavoring the emergence of an $N = 90$ shell closure~\cite{jungclaus2024}. The experiments~\cite{lozeva2016,lozeva2024} measured properties of Sb and Te nuclei close to $N=90$ in order to infer about the structure of $^{140}$Sn.  A definitive test of the presence of a shell closure would involve a direct measurement of the excitation energy of the first $2^+$ state in $^{140}$Sn. Nevertheless, experimental information on even Sn isotopes beyond $N = 82$ remains limited to $^{134,136,138}$Sn~\cite{korgul2000,simpson2014,piersasilkowska2021}, and $\gamma$-ray spectroscopy of $^{140}$Sn is currently beyond experimental reach.

In this work, we tackle this problem from first principles.  \textit{Ab initio} calculations  combine nuclear interactions derived from chiral effective field theory~\cite{weinberg1990,epelbaum2009,machleidt2011} with systematically improvable many-body methods~\cite{hagen2014,dickhoff2004,lu2019,hergert2016,stroberg2019}. The field has advanced rapidly in recent years, with calculations now reaching $^{208}$Pb~\cite{hu2022,hebeler2023,arthuis2024} and $^{266}$Pb~\cite{bonaiti2025}. Systematic investigation of ground-state properties along tin isotopes have also been performed~\cite{tichai2024,demol2026,hildenbrand2026}. Building on this progress, we compute the first $2^+$ excited state in $^{140}$Sn, and show that our results do not support the presence of a shell closure at $N = 90$.

This paper is structured as follows. Section~\ref{hamiltonian_sec} introduces the nuclear Hamiltonian employed throughout this work. Section~\ref{methods} provides a brief overview of the coupled-cluster and in-medium similarity renormalization group methods used in our calculations. In Section~\ref{sn133}, we validate the theoretical framework by studying the low-lying spectrum of $^{133}$Sn. In Section~\ref{sn140}, we present our results on the structure of $^{140}$Sn and draw conclusions in Section~\ref{concl}.

\section{Hamiltonian}\label{hamiltonian_sec}
The starting point of our calculations is the intrinsic nuclear Hamiltonian
\begin{equation}
    \hat{H} = \sum_{i<j} \left(\frac{(\mathbf{p}_i -\mathbf{p}_j)^2}{2mA} + \hat{V}^{NN}_{ij}\right) + \sum_{i<j<k} \hat{V}^{3N}_{ijk}. 
    \label{hamiltonian}
\end{equation}
It includes the intrinsic kinetic energy (i.e., the kinetic energy of the system when the center of mass is removed), and the nucleon-nucleon and three-nucleon potential. 
In this work we use the chiral force 1.8/2.0~(EM) from Ref.~\cite{hebeler2011}. It is obtained by applying similarity-renormalization-group~\cite{bogner2007} transformations to the nucleon-nucleon potential by \textcite{entem2003} and it includes the leading chiral three-nucleon forces~\cite{epelbaum2006}. The low-energy constants of the latter have been fitted on ground-state properties of nuclei with mass numbers $A = 3, 4$. 
This interaction provides accurate results for the ground-state energies and excitation spectra of nuclei across the nuclear chart, see, e.g., Refs.~\cite{stroberg2019,sun2025,heinz2025}. Most relevant for this work are the accurate reproductions of shell closures (using the energy of the first $J^\pi=2^+$ state as an indicator) in oxygen~\cite{simonis2016}, calcium, nickel~\cite{simonis2017}, and $^{208}$Pb~\cite{bonaiti2025}, and accurate predictions for $^{78}$Ni~\cite{hagen2017} and $^{100}$Sn~\cite{morris2018} before these were measured~\cite{taniuchi2019}.  Such benchmarks on ground- and excited-state energies across the nuclear chart make it an appropriate choice for the purposes of this paper.

\section{Methods}
\label{methods}
We solve the quantum many-body problem with the coupled-cluster method~\cite{coester1960,kuemmel1978,bartlett2007} in an angular-momentum-coupled scheme~\cite{hagen2008,hagen2014} and with the valence-space in-medium similarity renormalization group (VS-IMSRG)~\cite{tsukiyama2011,tsukiyama2012,hergert2016,stroberg2019,takayuki2020}. 

\subsection{Coupled-cluster theory}
In coupled-cluster theory, one starts from an exponential ansatz for the nuclear ground-state wave function 
\begin{equation}
    \ket{\Psi_0} = e^T \ket{\Phi_0}
\end{equation} 
where the reference state $\ket{\Phi_0}$ typically corresponds to a Hartree-Fock solution, and the cluster operator 
\begin{equation}
T = T_1 + T_2 + \dots + T_n+ \ldots + T_A
\end{equation}
is written as a sum of $n$-particle--$n$-hole ($n$p-$n$h) excitation operators. The most frequently used truncation is the coupled-cluster singles and doubles (CCSD) approximation, where $T=T_1+T_2$~\cite{bartlett2007}. This scheme captures about 90\% of the ground-state correlation energy in closed shell systems~\cite{bartlett2007,hagen2009b,sun2022}. 
The cluster amplitudes are chosen such that the reference state $|\Phi_0\rangle$
becomes an eigenstate of the similarity transformed Hamiltonian
\begin{equation}
\label{Hsim}
 \overline{H} = e^{-T} H e^T   \ .
\end{equation}
Higher accuracy can be reached with the CCSDT-1 framework~\cite{lee1984}, where the leading-order terms of the coupled-cluster singles, doubles and triples (CCSDT) approach~\cite{noga1987} are also included. 
We compute the ground-state energy of $^{140}$Sn with the CCSD approximation and also present CCSDT-1 results in not-too-large model spaces. 

To compute excited states of $^{140}$Sn, we use the equation-of-motion (EOM) method in the EOM-CCSD approximation~\cite{stanton1993}. In this approach one first computes the ground-state of $^{140}$Sn. Excited states are then computed as eigenstates of the similarity-transformed Hamiltonian, and one limits correlations  to one-particle-one-hole (1p-1h) and two-particle-two-hole (2p-2h) excitations. This is accurate for states that are dominated by 1p-1h excitations. 
To estimate the impact of higher-order many-body correlations, we also use the EOM-CCSDT-1 scheme~\cite{watts1995} for not-too-large model spaces. Due to computational resources, we need to restrict the number of 3p-3h excitations included in our calculations. Indicating with $\tilde{e}_p$ the difference in energy between the single-particle state $p$ and the Fermi surface, our EOM-CCSDT-1 computations include all 3p-3h configurations for which $\tilde{e}_{pqr} = \tilde{e}_p + \tilde{e}_r + \tilde{e}_q < 100$ MeV. 
 
As a validation of our description of shell structure in neutron-rich Sn isotopes, we also study the nucleus $^{133}$Sn where we can compare with data~\cite{jones2010,jones2011}. We compute its spectrum with the particle-attached EOM-CCSD (PA-EOM-CCSD) technique~\cite{gour2005,hagen2009b,hagen2010a}. Here, we first compute the ground-state of $^{132}$Sn with the coupled-cluster method, construct the corresponding similarity-transformed Hamiltonian~(\ref{Hsim}), and then compute low-lying states of $^{133}$Sn as eigenstates of $\overline{H}$ that consist of one-particle-zero-hole (1p-0h) and two-particle-one-hole (2p-1h) excitations. This is accurate for states with a dominant single-particle character. To assess the quality of this approximation, we compute perturbative three-particle-two-hole (3p-2h) corrections as in Ref.~\cite{morris2018}, which we denote as PA-EOM-CCSD(3p-2h)$_{\rm pert}$.  

\subsection{In-medium similarity renormalization group}
In the VS-IMSRG one applies a unitary transformation $U(s)$ to a given Hamiltonian $H$ to decouple the many-body problem into a smaller, more tractable space, the so-called valence space (VS). We have
\begin{equation}\label{eq:Hsdef}
    H(s) \equiv U(s) H U^{\dagger}(s) \ ,
\end{equation}
and the unitary transformation is parametrized by the flow parameter $s$, and  $U(0)=1$.

The goal is to construct a transformation such that, as $s$ increases, the off-diagonal part of the Hamiltonian is suppressed, $H^{\rm od}(s) \to 0$.
To obtain the desired transformation $U(s)$, one solves the flow equation for $H(s)$,
\begin{equation}\label{eq:dHds}
    \frac{d}{ds}H(s) = [\eta(s),H(s)],
\end{equation}
where $\eta(s)$ is the generator of the transformation.
The generator is required to be anti-Hermitian, $\eta^{\dagger}(s)=-\eta(s)$, and is chosen to depend on the off-diagonal part $H^{\rm od}(s)$.
The unitary transformation satisfies
\begin{equation}\label{eq:dUds}
    \frac{d}{ds}U(s) = \eta(s)U(s).
\end{equation}
There is some freedom in the choice of $\eta(s)$, and several forms have been used in the literature~\cite{glazek1993,wegner1994}. In this work, we use the arctangent variant of the White generator~\cite{white2002}.

The IMSRG transformation introduces many-body terms in both the Hamiltonian and other operators, which must be truncated because of their computational cost.
In this work, we adopt two VS truncation schemes, namely VS-IMSRG(2)~\cite{hergert2016} and VS-IMSRG(3f2)~\cite{stroberg2024}.
In IMSRG(2), all operators are truncated at the normal-ordered two-body level.
Since a straightforward extension to IMSRG(3)~\cite{heinz2021} is computationally too expensive, the IMSRG(3f2)~\cite{he2024} scheme is an approximation of the full IMSRG(3) that retains the three-body contributions scaling similar to IMSRG(2).
Then, the valence-space shell-model diagonalization is performed using the KSHELL code~\cite{Shimizu2019_KSHELL_BlockLanczos,Shimizu2013_KSHELL}.

As an alternative to solving the problem in the VS, we can directly decouple the reference state using the same unitary transformation technique. This provides us with a closed-shell ground state of $^{140}$Sn. 
To compute excited states of $^{140}$Sn, we can also use the EOM approximation within the IMSRG framework~\cite{parzuchowski2017b,parzuchowski2017}.
The numerical implementation employed the public IMSRG code~\cite{stroberg_code}, and a private code provided by Zhonghao Sun~\cite{Sun_private_code}.
The latter allows us to perform EOM-IMSRG(2) calculations with up to 2p-2h excitations for the first $2^+$ state of $^{140}$Sn.

\subsection{Model spaces}
Our calculations employ a Hartree-Fock single-particle basis, including single-particle excitations up to $N_{\rm max}\hbar\Omega$. For coupled-cluster calculations, Hartree-Fock computations starts from a standard shell-model filling of spherical harmonic oscillator orbitals for $^{132}$Sn and assume the presence of a 1$f_{7/2}$ neutron subshell closure in $^{140}$Sn. We employ model spaces consisting of up to 17 major harmonic oscillator shells, i.e., $N_{\rm max}=16$, and values of the harmonic oscillator frequency between  $\hbar\Omega = 10$ and 16~MeV. This allows us to assess the residual model space dependence of our results.

We use the codes~\cite{miyagi2023,stroberg_code} to compute the matrix elements of the three-body force, employing the normal-ordered two-body approximation~\cite{hagen2007a,roth2012,hebeler2023,rothman2025}, and apply an energy cut $E_{3,{\rm max}} = 28 \hbar\Omega$ on three-body matrix elements~\cite{takayuki2022}.
For the VS-IMSRG calculations, we take ${}^{132}$Sn as the core.
The valence space used in this work consists of the neutron $2p_{1/2}$, $2p_{3/2}$, $1f_{5/2}$, $1f_{7/2}$, $0h_{9/2}$, and $0i_{13/2}$ orbitals, while the proton shell is kept closed.
Because of computational limitations, we use up to 15 major harmonic-oscillator shells for VS-IMSRG(2) and 13 major shells for VS-IMSRG(3f2).

\section{Validation in $^{133}$S\NoCaseChange{n}}
\label{sn133}
For validation we compute the ground state and low-lying excited states of $^{133}$Sn, for which experimental data are available~\cite{urban1999,hoff1996,hoff2000,jones2010,dyszel2025}. 
The ground state of $^{133}$Sn has $J^{\pi} = 7/2^-$, and its ground-state energy amounts to $-1096$~MeV with the PA-EOM-CCSD(3p-2h)$_{\rm pert}$ approximation. The results obtained with the largest values of $N_{\rm max}$ ($N_{\rm max} = 12$ and $N_{\rm max} = 14$) differ by 7~MeV, less than $1\%$ of the total energy. The bulk of the ground-state energy is provided by the closed-shell reference, $^{132}$Sn. The PA-EOM-CCSD(2p-1h) computation lowers the total energy by approximately 1~MeV, and perturbative 3p-2h corrections amount to at most 0.4~MeV. For the IMSRG calculations, we use $N_{\rm max} = 12$.
The ground-state energy from the VS-IMSRG(2) and VS-IMSRG(3f2) schemes amounts to $-1106$ and $-1113$~MeV, respectively, around 1$\%$ lower than those from PA-EOM-CCSD. Both the coupled-cluster and IMSRG binding energies for the 1.8/2.0~(EM) interaction show a difference of less than $1\%$ from the experimental value of 1105.241837 MeV~\cite{ame2020}.

\begin{figure*}[htb]
    \centering
    \includegraphics[width=\textwidth]{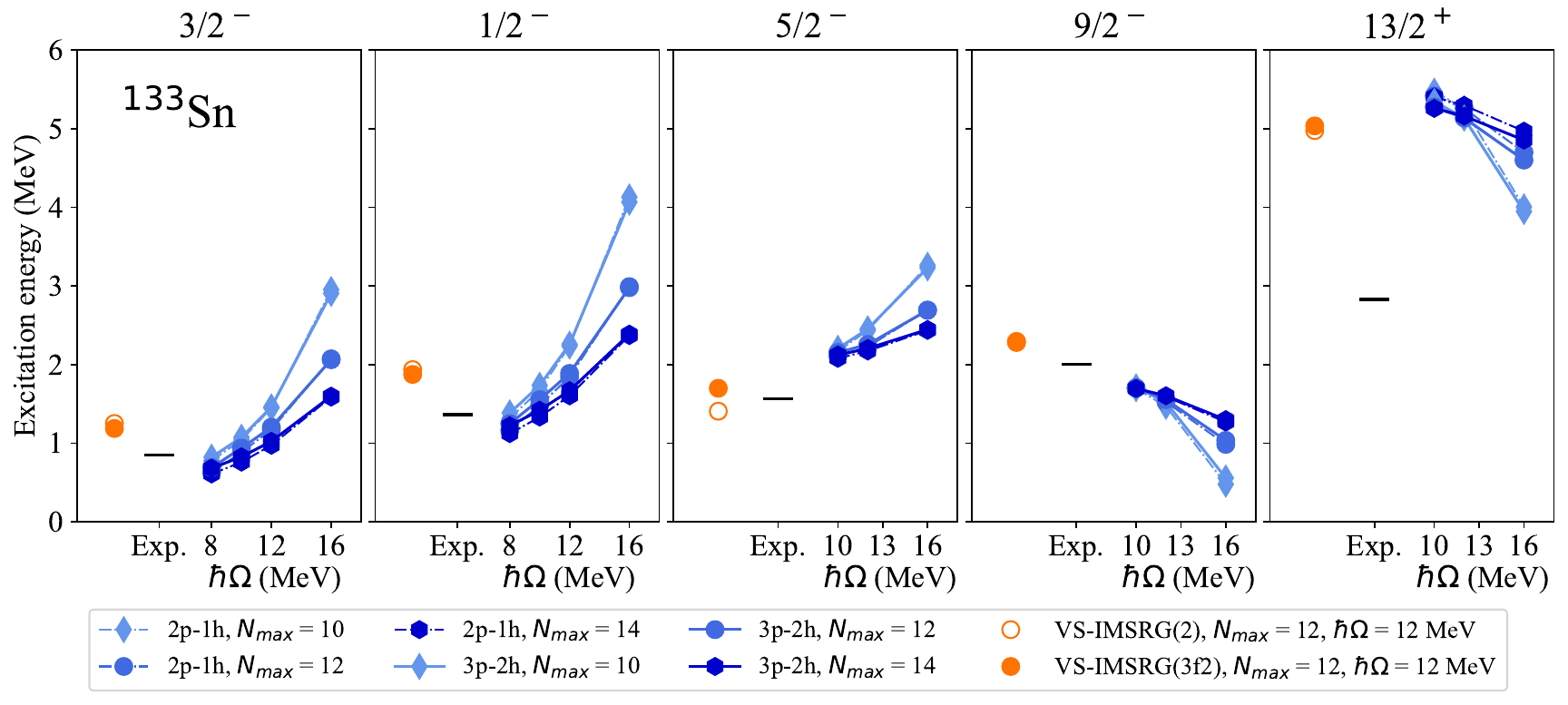}
    \caption{Model space convergence of the excited states of $^{133}$Sn, computed with the PA-EOM-CCSD(2p-1h) and PA-EOM-CCSD(3p-2h)$_{\rm pert}$ approaches employing the 1.8/2.0 (EM) interaction. For comparison, VS-IMSRG results obtained with $N_{\rm max} = 12$, $\hbar\Omega = 12$~MeV are reported, as well as experimental data from Refs.~\cite{urban1999,hoff2000,jones2010} for the $3/2^-$, $1/2^-$, $5/2^-$ and $9/2^-$ state and from Ref.~\cite{dyszel2025} for the $13/2^+$ state. }
    \label{fig:133Sn}
\end{figure*}

We also compute the low-lying $3/2^-$, $1/2^-$, $5/2^-$, $9/2^-$ and $13/2^+$ excited states of $^{133}$Sn. Excitation energies with respect to the $7/2^-$ ground state obtained with the 1.8/2.0~(EM) interactions are shown in Fig.~\ref{fig:133Sn} for both the coupled-cluster and VS-IMSRG approaches, in comparison to experimental data from Refs.~\cite{urban1999,hoff2000,jones2010,dyszel2025}. For coupled-cluster results, we also present the convergence pattern with respect to $N_{\rm max}$ and $\hbar\Omega$. 

For coupled-cluster results, a model space size of $N_{\rm max} = 14$ is sufficient to obtain converged results for the low-lying spectrum of $^{133}$Sn. At the optimal frequency, varying between $8$ and $12$~MeV depending on the state considered, variations between $N_{\rm max} = 14$ and $N_{\rm max} = 12$ amount to less than 0.1~MeV. Interestingly, the $3/2^-$, $1/2^-$ and $5/2^-$ converge from above, while the $9/2^-$ and $13/2^+$ exhibit the opposite pattern. This is due to the different optimal frequencies characterizing the PA-EOM-CCSD excitation energies of these states and of the $7/2^-$ ground state. We surmise that an explanation could be based on orbital angular-momentum barriers. The 1p-0h contribution to the wavefunction's norm is around $90\%$ for all the computed states in $^{133}$Sn. This suggests that such states are of single-particle character, and it is consistent with the small effect of perturbative 3p-2h corrections on their excitation energy. 

We also performed VS-IMSRG(2) and VS-IMSRG(3f2) computations for the spectrum of $^{133}$Sn. 
Results are overall consistent with coupled-cluster calculations. 
The inclusion of double-commutator corrections within the VS-IMSRG(3f2) framework for $^{133}$Sn yields only small corrections relative to the VS-IMSRG(2) result.
The single-particle energies of $^{133}$Sn obtained with VS-IMSRG(3f2) are almost unchanged compared with those obtained with IMSRG(2). 
This is expected because, near the shell closure, the spherical reference is a very good approximation, and IMSRG(2) already captures the dominant contributions. 
The three-body correlations provide only tiny corrections.

Overall, PA-EOM-CCSD and VS-IMSRG results are in agreement with data for the $3/2^-$, $1/2^-$, $5/2^-$ and $9/2^-$ states, while they both overestimate the excitation energy of the $13/2^+$ state by a factor of about two. This state was only precisely determined precisely most recently via $\beta$-delayed neutron spectroscopy~\cite{dyszel2025}. Recently, it has also been studied via the $^{132}$Sn(d,p) reaction~\cite{kay_priv}. We are not sure why theory obtains this state so much higher than observed. On the one hand, this state is neutron-unbound and the lack of continuum in our calculations could play a role. On the other hand, the formidable $l=6$ angular momentum barrier will reduce its impact. It would be interesting to explore the Hamiltonian dependence of this state. This task is left for future work.

\section{Structure of $^{140}$S\NoCaseChange{n}}
\label{sn140}
 
\begin{figure}[htb]
    \centering
    \includegraphics[width=0.49\textwidth]{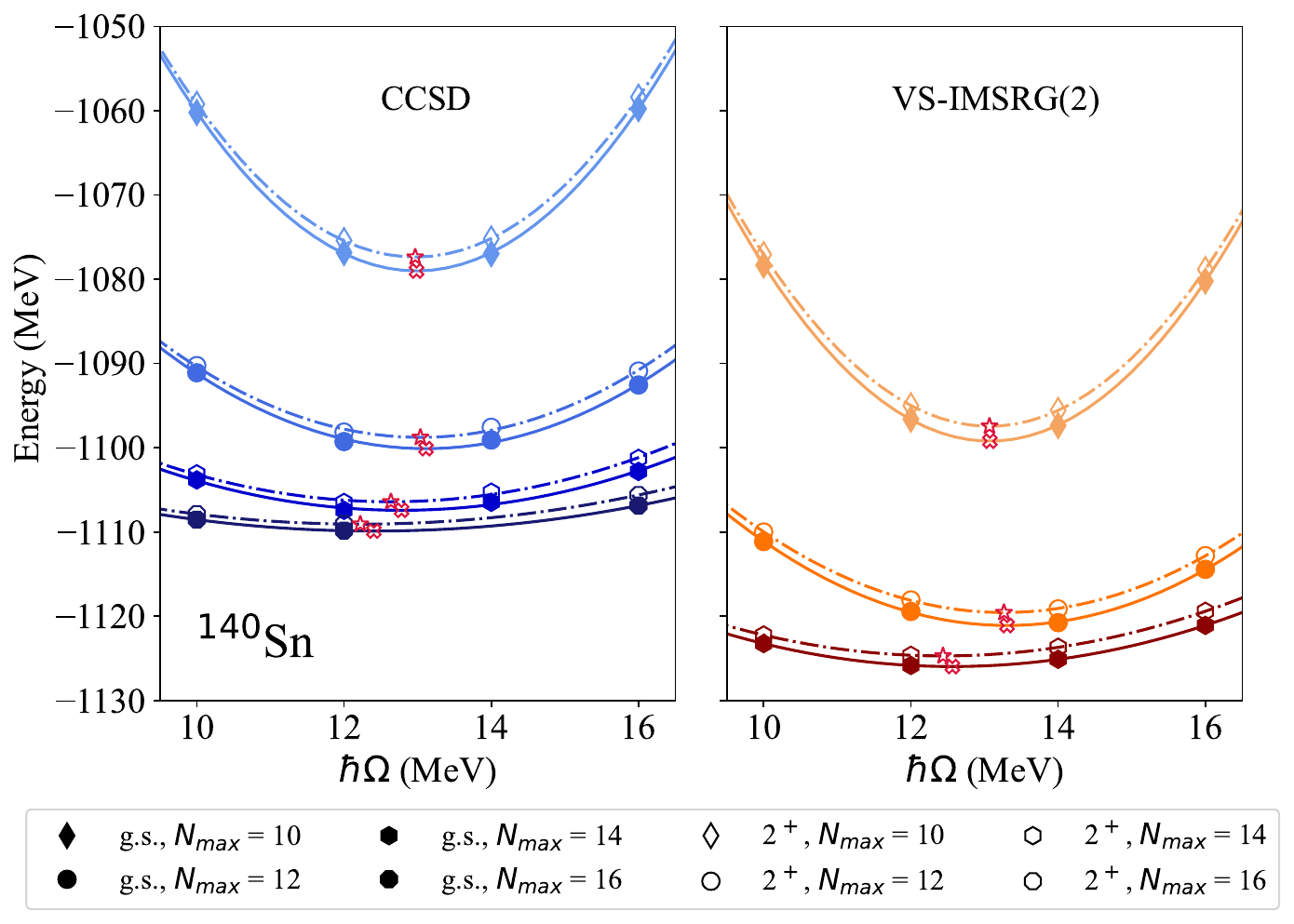}
    \caption{Model space convergence of the total ground-state and first excited $2^+$ state energy calculated with the CCSD and VS-IMSRG(2) approaches employing the 1.8/2.0 (EM) interaction. Solid (dashed) lines and full (empty) markers are results for the ground state ($2^+$ state). The red crosses (stars) denote the minimum points determined by a parabolic fit to the numerical data of the ground state ($2^+$ state). }
    \label{fig:140Sn_conv}
\end{figure}

Figure~\ref{fig:140Sn_conv} shows how the energies of the ground-state  and  the first $2^+$ state converge as the size of the model space is increased. For coupled-cluster computations, the CCSD approximation is employed for the ground-state, while the first $2^+$ state is computed at the EOM-CCSD level. In this case, we were also able to obtain a limited set of results with $N_{\rm max} = 16$. For VS-IMSRG computations, the VS-IMSRG(2) approximation is adopted for both states. 

We observe that optimal oscillator frequencies for the ground state energy are a bit larger than those of the $2^+$ state. Let us first focus on the ground state. Taking the minima of the parabolic curves at the highest $N_{\rm max}$ available ($N_{\rm max} = 16$ for CCSD and $N_{\rm max} = 14$ for VS-IMSRG(2)), we find a ground-state energy of $-1109$~MeV for CCSD and of $-1125$~MeV for VS-IMSRG(2), with a residual dependence on the model space size below $1\%$ for both approaches. We notice that CCSD yields a higher ground-state energy than VS-IMSRG(2). This effect is related to a different counting of many-body contributions in the two methods~\cite{hergert2016}.  

While the binding energy of $^{140}$Sn has not been measured, trends of the mass surface estimate it to be at $-1125$~MeV~\cite{ame2020}. The FRDM(2012) mass model~\cite{moller2012} gives a similar prediction of $-1124.43$ MeV. For small model spaces ($N_{\rm max} = 6, 8$), we can perform CCSDT-1 calculations. The latter confirm the observation according to which triples contributions typically amount to $10\%$ of the correlation energy. Under this assumption, starting from the CCSD prediction for $N_{\rm max} = 16$, coupled-cluster calculations with the 1.8/2.0~(EM) interaction would overbind $^{140}$Sn with respect to nuclear mass models of $1\%$. 
The prediction from IMSRG(3f2) with $N_{\rm max} = 12$ is $-1126$~MeV, which is remarkably close to mass model estimates.
However, we note that the IMSRG(3f2) result is not fully converged at $N_{\rm max} = 12$. 
For $^{140}$Sn, the IMSRG(3f2) calculation yields a slightly more bound ground state than IMSRG(2). 
The IMSRG(3) correction to the ground-state energy relative to IMSRG(2) is not universal. 
It depends on the competition between the additional triples correction and the factorized double-commutator contributions specific to $^{140}$Sn in IMSRG(3f2)~\cite{he2024}.

\begin{figure}[htb]
    \centering
    \includegraphics[width=0.49\textwidth]{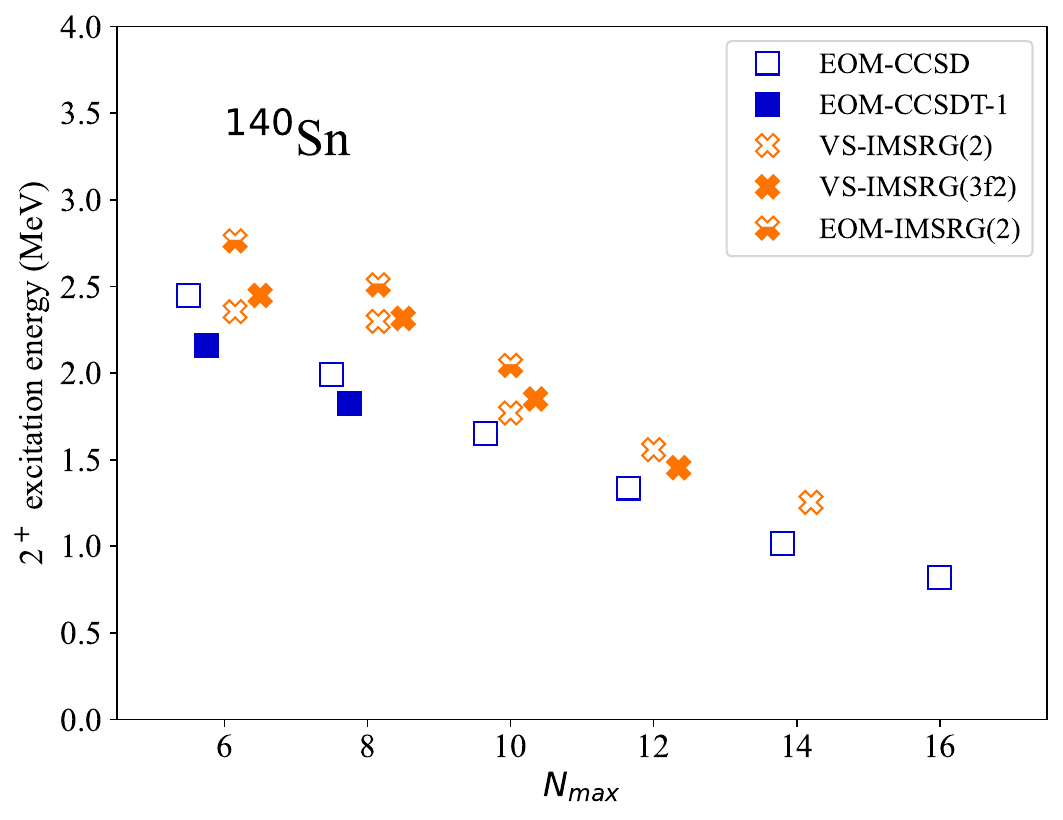}
    \caption{Convergence of the first $2^+$ state excitation energy as a function of the model-space size $N_{\rm max}$ for the 1.8/2.0 interaction, computed with the EOM-CCSD, EOM-CCSDT-1, VS-IMSRG(2), VS-IMSRG(3f2) and EOM-IMSRG(2) approaches.}
    \label{fig:140Sn_conv1}
\end{figure}

We plot the excitation energy of the first $2^+$ state, i.e. the difference between the minimal total energies of the $2^+_1$ and $0^+_1$ states, at fixed $N_{\rm max}$ in Fig.~\ref{fig:140Sn_conv1}. For an estimate of our many-body uncertainty, we include EOM-CCSDT-1 results obtained with $\tilde{e}_{pqr} < 100$ MeV for smaller model spaces, as well as VS-IMSRG(3f2) and EOM-IMSRG(2) results.  As observed previously in  Refs.~\cite{hagen2017,morris2018,bonaiti2025}, the inclusion of 3p-3h correlations reduces the excitation energy by up to  $0.5$~MeV in this case. The results are not yet converged with increasing $N_\mathrm{max}$ (but decrease monotonically). This indicates the absence of a shell gap in $^{140}$Sn and is the main result of this paper. 

The absence of a shell gap is qualitatively supported by  EOM-CCSD(T)~\cite{watts1995} calculations, where 3p-3h corrections are included perturbatively. This approach yields negative excitation energies, indicating a complete breakdown of the approach. This indicates that the basic assumption, i.e. a closed-shell reference state, is wrong. This can be understood by looking at the extent of the shell gap at the mean-field level. At $N_{\rm max} = 16$, our computations show that the neutron shell gap is only about $2$~MeV, while the corresponding proton shell gap is approximately $7$~MeV. This already suggests that the reference state is not separated by a robust neutron shell gap at $N=90$.

In the IMSRG computations, valence-space results are lower than those from EOM-IMSRG(2). The addition of flowing three-body correction in valence-space calculations slightly increases the excitation energy for $N_{\rm max} = 6, 8, 10$, while we observe the opposite behaviour for $N_{\rm max} = 12$. In this case, it is worth pointing out that VS-IMSRG(3f2) results for $N_{\rm max} = 12$ could be obtained only with $E_{3,\rm max} = 24$. 
Compared to the EOM-IMSRG(2) calculation for the $2^+$ state, which can only include up to 2p-2h excitations, a VS treatment generally includes more many-body correlations than EOM-IMSRG at the same truncation level. This might explain the lower $2^+$ state obtained in the VS framework. The difference between VS-IMSRG and EOM-IMSRG results can thus be treated as an estimate of the IMSRG many-body uncertainty. 

IMSRG results lie slightly higher than those from coupled-cluster theory, but in both approaches, the $2^+$ excitation energy in $^{140}$Sn is characterized by a slow convergence pattern. Our predictions, based on the 1.8/2.0 (EM) interaction, clearly indicate that the $2^+$ excitation energy is lower than 1.5~MeV. This value is less than $60\%$ of the first $2^+$ excitation energy of the most neutron-rich closed-shell nucleus $^{78}$Ni~\cite{taniuchi2019}. Doubly-magic nuclei as $^{40,48}$Ca and $^{208}$Pb have  excitation energies that are more than $50\%$ larger~\cite{nudat}. At $N_{\rm max} = 16$, coupled-cluster calculations yield a $2^+$ excitation energy below $1$~MeV. This allows us to exclude the possibility of a doubly-magic shell closure in $^{140}$Sn. Our findings are consistent with previous shell model calculations, predicting the $2^+$ excitation energy in $^{140}$Sn  between $0.5$ and $1$~MeV~\cite{covello2011,naidja2015}. Such a conclusion is also supported by recent \textit{ab initio} investigations of the structure of tin isotopes, based on Bogoliubov coupled-cluster theory, where no clear gap is detected in two-neutron separation energies~\cite{tichai2024,vernik2026}, and no kink is found in trends of charge radii~\cite{demol2026} at $N = 90$.

\section{Summary}
\label{concl}
We computed the low-lying states of $^{133}$Sn and $^{140}$Sn, employing the 1.8/2.0 (EM) interaction and two different \textit{ab initio} methods. Excited states in $^{133}$Sn are in good agreement with experimental data, except for the case of the neutron-unbound $13/2^+$ state. Our predictions for the ground-state energy of $^{140}$Sn are consistent with estimates from nuclear mass models. We find a small excitation energy for the first $2^+$ state, which does not support the hypothesis of a (sub)shell closure in $^{140}$Sn.  

\section*{Data availability}
The data that support the findings of this article are openly available~\cite{bonaiti_sn140_zenodo2026}.

\begin{acknowledgments}
We thank Matthias Heinz for useful comments on the manuscript. B.C.H. thanks Zhonghao Sun for providing the EOM-IMSRG code. This work was supported by the U.S. Department of Energy, Office of Science, Office of Nuclear Physics, under the FRIB Theory Alliance award DE-SC0013617 and award No.~DE-FG02-96ER40963; by the U.S. Department of Energy, Office of Science, Office of Advanced Scientific Computing Research and Office of Nuclear Physics, Scientific Discovery through Advanced Computing (SciDAC) program (SciDAC-5 NUCLEI); by the National Science Foundation via the Focused Research Hubs in Theoretical Physics program under award No.~PHY-2402275.
This research used resources of the Oak Ridge Leadership Computing Facility located at Oak Ridge National Laboratory, which is supported by the Office of Science of the Department of Energy under contract No. DE-AC05-00OR22725. Computer time was provided by the Innovative and Novel Computational Impact on Theory and Experiment (INCITE) program. 
This research was supported in part by the Notre Dame Center for Research Computing through the Notre Dame Research Computing Cluster.
\end{acknowledgments}

\bibliography{master}

\end{document}